\documentclass[11pt]{article}

\usepackage{amsmath}
\usepackage{graphicx}
\usepackage{amsfonts}
\usepackage{amssymb}
\usepackage{epsfig}
\usepackage{color}
\usepackage{psfrag}

\newcommand{\EQ}{\begin{equation}}
\newcommand{\EN}{\end{equation}}
\newcommand{\be}{\begin{equation}}
\newcommand{\ee}{\end{equation}}
\newcommand{\bea}{\begin{eqnarray}}
\newcommand{\eea}{\end{eqnarray}}

\newcommand{\goto}{\rightarrow}

\def\goto{\longrightarrow}
                                                                                                                               
\begin{document}

\setcounter{page}{0} \topmargin0pt \oddsidemargin5mm \renewcommand{%
\thefootnote}{\arabic{footnote}}\newpage \setcounter{page}{0} 
\begin{titlepage}
\begin{flushright}
SISSA 08/2009/EP
\end{flushright}
\vspace{0.5cm}
\begin{center}
{\large{\bf Field theory of Ising percolating clusters}
}\\

\vspace{1.8cm}
{\large Gesualdo Delfino
} 
\\
\vspace{0.5cm}
{\em International School for Advanced Studies (SISSA)}\\
{\em via Beirut 2-4, 34014 Trieste, Italy}\\
{\em INFN sezione di Trieste}\\
\end{center}
\vspace{1.2cm}

\renewcommand{\thefootnote}{\arabic{footnote}}
\setcounter{footnote}{0}

\begin{abstract}
\noindent
The clusters of up spins of a two-dimensional Ising ferromagnet undergo a 
second order percolative transition at temperatures above the Curie point. We 
show that in the scaling limit the percolation threshold is described by an 
integrable field theory and identify the non-perturbative mechanism 
which allows the percolative transition in absence of thermodynamic 
singularities. The analysis is extended to the Kert\'esz line along which the 
Coniglio-Klein droplets percolate in a positive magnetic field.
\end{abstract}

\end{titlepage}

\newpage

\section{Introduction}
The Ising model provides the fundamental example of critical behavior 
produced by the interaction  among an infinite number of degrees of freedom,
represented by spins on a  lattice. Another type of criticality, which requires no
interaction and, for this reason, is often called geometrical, finds its simplest
illustration in random percolation \cite{SA}. Here the sites of the lattice are
randomly occupied
 with probability $p$, and criticality (the appearance of an infinite cluster of 
nearest-neighbor occupied sites) arises when $p$ reaches a lattice-dependent
value $p_c$. 

On the other hand, a ferromagnetic system like the Ising model encodes itself a
percolation problem: nearest-neighbor sites with spin `up' form clusters which 
can percolate for specific values of the parameters (temperature $T$ and magnetic
field $H$) which determine the interaction among the spins. This kind of percolation
problem, which is completely determined by the magnetic properties and does not 
affect them, is called correlated percolation. 
The comparison of the magnetic and percolative phase diagrams on the $T$-$H$
plane is particularly interesting. It turns out that in two dimensions (not in three) the
temperature $T_p$ at which clusters percolate in zero field coincides with the 
Curie temperature $T_c$ \cite{CNPR}. Even in this case, however,  a 
second order percolation line goes from $(T,H)=(T_c,0)$ to a finite value 
of the magnetic field at infinite temperature, a circumstance which can appear
paradoxical if one considers that there are no thermodynamic singularities 
above $T_c$ and that spin-spin correlations decay exponentially.

Essential insight into this problem comes from the Kasteleyn-Fortuin 
representation \cite{KF} which associates the percolative properties to an 
auxiliary site variable $s_i$ which takes $q$ values. As a result,  the 
magnetic and percolative transitions, being related to different site 
variables, are in a sense disentangled. The price to pay is that $s_i$, 
being auxiliary, cannot remain in the game to the end: everything 
needs to be evaluated in the limit $q\to 1$ in which the Ising spins 
are the only real degrees of freedom. The solution of the above 
paradox then requires to deal with the subtleties of this limit. 

Field theory, as the natural framework for dealing with second order
critical points and the scaling region around them, is in principle 
the right place where to address this problem and isolate its universal
features. The difficulty, however,
is in the fact that the answer can hardly come from a perturbative
approach. In this paper we show that the percolation line above $T_c$
in two dimensions actually corresponds, in the scaling limit, to an
integrable field theory. On its exact solution, we can perform the 
limit $q\to 1$ analytically, unveiling the presence of $q-1$ massless
particles together with a massive unstable particle with lifetime
inversely proportional to $q-1$. Percolative properties are determined
at first order in $q-1$, where the theory is massless and describes the
crossover from the correlated to the random percolation fixed point. 
The magnetic properties are determined instead at $q=1$, where
there are no massless excitations left and the massive particle has 
become stable, providing the required finite correlation length.

Most of this discussion for the clusters can be repeated
for the Coniglio-Klein `droplets'  \cite{CK} which, for $H=0$ (also 
in three dimensions), satisfy the requirement of Fisher's droplet model 
\cite{Fisherdroplet}: $T_p=T_c$ and percolative exponents equal to the magnetic
ones. Also the Coniglio-Klein droplets exhibit for $H>0$ a second 
order percolation line, known as the Kert\'esz line \cite{Kertesz}, going 
this time from the Curie point to a finite value of the temperature at 
infinite magnetic field. In the scaling limit this line corresponds again
to a renormalization group trajectory within the Ising field theory, 
which, however, in this case is not integrable, so that its origin through
the resonance mechanism observed for the clusters, although very likely, 
cannot be followed analytically.

The paper is organized as follows. In the next section we review the
main results about the phase diagram of Ising clusters before 
recalling in section~3 the Kasteleyn-Fortuin formulation of the 
problem and the renormalization group analysis which leads to 
the identification of two different fixed points for clusters and droplets. 
The field theoretical discussion of clusters and droplets is then 
presented in sections~4 and 5, respectively, while section~6 contains a 
summary of the main conclusions.

\section{Percolation of Ising clusters}
Consider the ferromagnetic Ising model defined by the reduced 
Hamiltonian
\EQ
-{\cal H}_{Ising}=\frac1T\sum_{\langle ij\rangle}\sigma_i\sigma_j+H\sum_i
\sigma_i\,,\hspace{1cm}\sigma_i=\pm 1\,,
\label{ising}
\EN
where $\sigma_i$ is a spin variable located at the $i$-th site of an infinite 
regular lattice, $T\geq 0$ and $H$ are couplings that we call temperature and 
magnetic field, respectively, and the first sum is restricted to 
nearest-neighbor spins. For $H=0$ the model is well known to exhibit a non-zero
magnetization per site $M=\langle\sigma_i\rangle$ at temperatures below a 
critical value $T_c$ (the magnetization in zero field is called spontaneous). 
The magnetization has a discontinuity at $H=0$ along a path taken at 
fixed $T<T_c$ on the $T$-$H$--plane (first order transition), and vanishes when
$T_c$ is approached from below at $H=0$ (second order transition). No other
magnetic transition (i.e. discontinuity in $M$ or its derivatives) takes place
away from $T\leq T_c$, $H=0$.

If, given a spin configuration, we draw a link between nearest-neighbor `up'
spins (i.e. spins taking the value $+1$), we obtain connected sets of up spins
that we call clusters. Of course, clusters of `down' ($-1$) spins are
defined analogously; in the following, talking of clusters without further 
specification we will refer to clusters of up spins, being understood that 
similar statements hold for the clusters of down spins under the 
substitution $H\goto-H$.

For $H=+\infty$ all the spins are forced to be up, so that the whole lattice is
occupied by a unique infinite cluster. The fraction $P$ of the lattice 
occupied by this infinite cluster decreases when the magnetic field decreases 
at fixed $T$, and is certainly zero at $H=-\infty$. We denote by $H_0(T)$ the 
value of the magnetic field below which the infinite cluster is absent. Since 
at $T=0$ all the spins are up for $H=0^+$, we have $H_0(0)=0^+$. When $T=
\infty$, on the other hand, the spins are uncorrelated and take the value $+1$
with probability $e^H/2\cosh H$; hence, denoting by $p_c^0$ the percolation
threshold for the random site percolation problem, we have
\EQ
\frac{e^{H_0(\infty)}}{2\cosh H_0(\infty)}=p_c^0\,.
\label{pc0}
\EN
Like $T_c$, $p_c^0$ is non-universal, i.e. depends on the structure of the 
lattice. Notice for example that, if $p_c^0<1/2$, $H_0(\infty)$ is negative,
so that at sufficiently high temperature an infinite cluster of up spins 
exists also in a negative field and coexists with an infinite cluster of down
spins in an interval of $H$ around zero.

For finite temperatures the Ising model encodes a generalized percolation 
problem in which the sites are not independent but interact through the 
Hamiltonian (\ref{ising}). The probability that a site has spin up is $(M+1)/2$
in terms of the average magnetization per site. Some early studies on Ising 
clusters can be found in \cite{Domb,M-K,BSM,Coniglio75,OOM,CNPR,SG}. Here we 
summarize the main evidences relying on the following three statements:

{i)} $H_0(T)$ is a monotonic function;

{ii)} the existence of a spontaneous magnetization implies the presence of
an infinite cluster;

{iii)} let $p$ be the probability that a site has spin up at $H=0$,
and $p_c$ the critical value above which an infinite cluster appears. Then
$p_c$ does not exceed the value $p_c^0$ of random percolation.

Statement {i)} can be justified observing that, as a consequence of 
the ferromagnetic interaction, the fraction of up (down) spins increases as we
decrease the temperature for a fixed positive (negative) value of the magnetic
field, and eventually becomes 1 at $T=0$. Hence, it is reasonable to expect 
that, if $H_0(\infty)$ is positive (negative), the strenght of the positive 
(negative) field needed to produce (destroy) the infinite cluster of up spins 
decreases with the temperature, until it vanishes (it is zero at $T=0$). 

Statement {ii)} is a rigorous result of \cite{CNPR}. Notice that the 
opposite is not true: for $H_0(\infty)<0$ there is an infinite cluster at 
large enough temperatures in zero field, but there is no spontaneous 
magnetization above $T_c$. Also, magnetization at $H\neq 0$ does not imply an
infinite cluster: if $0<H<H_0(\infty)$, there is no infinite cluster at 
infinite temperature in spite of the positive magnetization.

Statement iii) is an observation of \cite{Coniglio75}. A way of understanding 
it is to think of the ferromagnetic interaction as inducing an attraction 
among the clusters, which favors (with respect to the random case)
the formation of larger clusters, and eventually of the infinite cluster. 
Indeed, if $Z_{Ising}$ is the partition function, the probability of a 
configuration is $e^{-{\cal H}_{Ising}}/Z_{Ising}$. While configurations with a
fixed number of up spins are all equally probable in the random case, in 
presence of the interaction cluster formation lowers the energy and increases 
the probability.

We can now distinguish three cases according to the value of the random 
percolation threshold $p_c^0$:

a) $p_c^0>1/2$, i.e. $H_0(\infty)>0$. In this case i) prevents the coexistence
of infinite clusters of up spins with infinite clusters of down spins. $H_0(T)$
takes the value $0^+$ at $T=0$ and is forced by ii) to stay constant up to
$T_c$; above $T_c$ it is free to increase and to reach its positive asymptotic
value (Fig.~1). So $T_c$ is also the percolation point in zero field; since 
the magnetization vanishes at this point, one has $p_c=1/2$, which is 
consistent with iii).

b) $p_c^0<1/2$, i.e. $H_0(\infty)<0$. In this case $H_0(T)$ is zero up to some
value $T_p$ of the temperature, above which it becomes negative, so that an 
infinite cluster of up spins coexists with an infinite cluster of down spins
in the region $|H|<-H_0(T)$. At $H=0^-$, the fraction $p$ of sites with spin 
up, which is zero at $T=0$, increases with the temperature and reaches the
critical value $p_c$ for $T=T_p$. Since $p=1/2$ for $T\geq T_c$, iii) requires 
$T_p<T_c$. Hence, for $T_p<T<T_c$ there are infinite clusters, with different 
origin and different density, on both sides of $H=0$.

c) $p_c^0=1/2$, i.e. $H_0(\infty)=0$. In this case $H_0(T)=0$ at all 
temperatures. Continuity with the previous two cases requires that $T_p\to 
T_c^-$ as $H_0(\infty)\to 0^-$, so that, like in case a), the percolative 
transition is first order below $T_c$ and, presumably, second order above.

This classification according to $p_c^0$ leads to a distinction between the 
two- and three-dimensional cases \cite{Coniglio75}. Indeed, the two-dimensional
lattices have\footnote{In a given dimension $d$, the critical probability 
$p_c^0$ of random site percolation decreases as the coordination number $C$ 
of the lattice increases. In $d=2$, the triangular lattice ($C=6$) has 
$p_c^0=1/2$; in $d=3$, the diamond lattice ($C=4$) has $p_c^0\simeq 0.43$ 
(see \cite{SA} ).} $p_c^0\geq 1/2$ \cite{Harris,Fisher}, 
and then fall into the cases a), c), which share the same critical pattern
with both the magnetic and the percolative transition taking place at $T_c$ 
in zero field. For the square lattice Ising model this was confirmed by
series expansions in \cite{SG} and proved rigorously in \cite{CNPR}. 
For three-dimensional lattices the evidence
is that $p_c^0<1/2$ \cite{Essam,SGG}, so that they fall into the case b), for 
which the magnetic and percolative transitions in zero field are not 
simultaneous. For the simple cubic lattice this was first seen in \cite{M-K} 
using Monte Carlo simulations. 

\begin{figure}
\centerline{
\includegraphics[width=7cm]{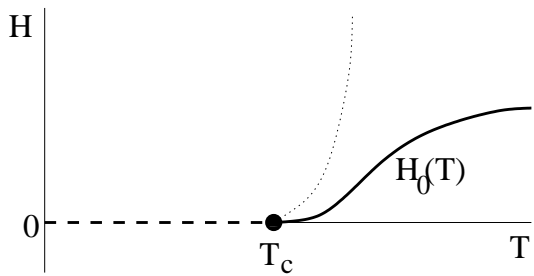}}
\caption{Qualitative phase diagram expected when the critical probability
$p_c^0$ of random site percolation is larger than $1/2$. The thick line is 
the curve $H_0(T)$ above which there is an infinite cluster of up spins. Its
dashed portion indicates that the percolative transition is first order below 
$T_c$. The Kert\'esz line (dotted) will be discussed in section~5.
}
\end{figure}

\section{Kasteleyn-Fortuin representation}
\subsection{Dilute Potts model}
The correlated percolation problem introduced in the previous section admits
a formulation in terms of auxiliary Potts variables which generalizes that 
originally given for random percolation by Kasteleyn and Fortuin \cite{KF}. 

Let us first of all rewrite, up to an inessential additive constant, the 
Hamiltonian (\ref{ising}) in the lattice gas language
\EQ
-{\cal H}_{Ising}=\frac4T\sum_{\langle ij\rangle}t_it_j+\Delta\sum_it_i\,,
\hspace{1cm}t_i=(\sigma_i+1)/2=0,1\,,
\label{lg}
\EN
where $\Delta=2H-a/T$, with $a$ a lattice-dependent constant.
Then consider the dilute $q$-state Potts model with Hamiltonian
\EQ
-{\cal H}_q=-{\cal H}_{Ising}+J\sum_{\langle ij\rangle}t_it_j\,(\delta_{s_i,
s_j}-1)\,,\hspace{1cm}s_i=1,2,\ldots,q\,.
\label{potts}
\EN
The partition function allows the following representation in which the 
sum over the Potts variables $s_i$ is replaced by one over bond variables  
\cite{Murata}
\bea
Z_q &=& \sum_{\{t_i\}}\sum_{\{s_i\}}e^{-{\cal H}_q}\nonumber\\
  &=& \sum_{\{t_i\}}e^{-{\cal H}_{Ising}}\,q^{N_v}\sum_G p_B^b
(1-p_B)^{\bar{b}}\,q^{N_c}\,,\hspace{1cm}p_B\equiv 1-e^{-J}\,,
\label{zq}
\eea
where $N_v$ is the number of empty ($t_i=0$) sites, the last sum is performed
over the graphs $G$ obtained putting bonds in all possible ways between 
nearest-neighbor sites belonging to the restricted lattice formed by the 
occupied ($t_i=1$) sites only, $b$ is the number of bonds in the graph $G$, 
$\bar{b}$ is the number of absent bonds on the restricted lattice, and $N_c$ is
the number of connected components in $G$. Such connected components are called
Kasteleyn-Fortuin (KF) clusters (Fig.~2). 

\begin{figure}
\centerline{
\includegraphics[width=4cm]{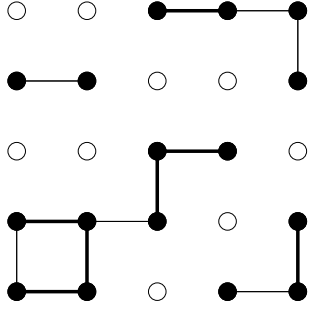}}
\caption{Site-bond configuration on the square lattice. The full (empty) 
circles correspond to occupied (empty) sites of the lattice gas. Bonds (thick
segments) between nearest-neighbor occupied sites are present with probability 
$p_B$ and define the KF clusters. For $p_B=1$ all segments are thick and the 
KF clusters coincide wiht the Ising clusters.  
}
\end{figure}

For $q=1$ there are no Potts degrees of freedom ($Z_1=Z_{Ising}$) and the 
second sum in (\ref{zq}) gives 1, showing that $p_B$ is simply the probability 
that a bond is present on the restricted lattice. By construction, the KF 
clusters live on the clusters of up spins of the Ising model, and become the 
Ising clusters for $p_B=1$.

The number $q$ of Potts colors appears in (\ref{zq}) as a
parameter (which can be taken continuous) through power-like terms accounting
for the fact that both the empty sites and the KF clusters can take $q$ colors.
If $X$ is an observable, the average associated to the partition function
(\ref{zq}) with $q=1$ is
\EQ
\langle X\rangle=Z_{Ising}^{-1}
\sum_{\{t_i\}}e^{-{\cal H}_{Ising}}\sum_G X\,p_B^b(1-p_B)^{\bar{b}}\,. 
\label{average}
\EN
When $H=+\infty$ there are no vacancies and the average reduces to the 
weighted sum over bond configurations on the whole lattice, i.e. the usual
random bond percolation problem. For generic values of $T$ and $H$, 
instead, (\ref{average}) corresponds to a generalized percolation problem in 
precence of vacancies whose distribution is weighted by the Ising lattice gas
Hamiltonian. For example, for $X=N_c$, (\ref{average}) gives the mean cluster
number in this generalized percolation problem, as a function of $p_B$. 
Evaluated at $p_B=1$, this quantity is the mean number of spin up clusters in
the Ising model at the given values of $T$ and $H$.

Notice that, while the correlations among Ising spins affect the associated
bond percolation problem, the opposite is not true. At $q=1$ the Hamiltonian
(\ref{potts}) reduces to ${\cal H}_{Ising}$ and the parameter $J$ (or $p_B$)
plays no dynamical role: the conjugated bond variables over which the second
sum in (\ref{average}) is performed only serve enumeration purposes. In 
particular, an observable $X$ which does not depend on the bond variables goes
out of the second sum, which then gives 1, leaving us with the usual Ising
thermodynamic average.

The role of the parameter $q$ is further clarified if we consider the 
Hamiltonian $\tilde{\cal H}_q$ obtained adding to ${\cal H}_q$ the term
$\tilde{H}\sum_it_i(\delta_{s_i,1}-1)$, where the magnetic field $\tilde{H}$
conjugated to the Potts variable is usually called `ghost' field. The new
partition function $\tilde{Z}_q$ is then given \cite{Murata} by $Z_q$ with 
$q^{N_c}$ replaced by $\prod_r\left[(q-1)e^{\tilde{H}S_r}+1\right]$, where $r$ 
labels the KF clusters in $G$ and $S_r$ is the number of sites in the $r$-th 
cluster. When we expand the free energy per site around $q=1$,
\EQ
\tilde{f}_q=-\frac1N\ln\tilde{Z}_q=f_{Ising}-(q-1)F+O((q-1)^2)\,,
\label{free-energy}
\EN
the function $dF/d\tilde{H}$ depends on $\tilde{H}$ through terms containing 
the factor $1/N\sum_rS_re^{\tilde{H}S_r}=\sum_Sn_SSe^{\tilde{H}S}$, where $n_S$
is the number of clusters of size $S$ per site, so that
\EQ
\left(\frac{d^kF}{d\tilde{H}^k}\right)_{\tilde{H}=0}=\sum_SS^k\langle n_S
\rangle\,,
\label{moments}
\EN
i.e. $F$ is the generating function for the moments of the cluster size 
distribution. The probability that a set of sites belongs to the same cluster
can be otained along the same lines introducing a site-dependent ghost field
\cite{Murata}. The important message of (\ref{free-energy}) and (\ref{moments})
is that the dilute $q$-state Potts model coincides with the 
Ising model at $q=1$ and determines the properties of KF clusters within the 
lattice gas at first order in $q-1$; it describes Ising clusters at first
order in $q-1$ when $p_B=1$.

\subsection{Renormalization group analysis}
The Hamiltonian description in terms of the dilute Potts model allows a
renormalization group analysis of percolative properties within the lattice 
gas. The case we are interested in is the two-dimensional one, for which
the magnetic and percolative transitions in the Ising model both take 
place at $T_c$ for $H=0$. The renormalization group analysis for this case was 
performed in \cite{CK} (see also \cite{CP,SV}). We now recall the main 
results of this analysis casting them within the field theoretical 
language needed for our subsequent purposes.

We look for fixed points of the Hamiltonian (\ref{potts}) in $d=2$ for $q\to 
1$. Since the percolative properties do not affect the magnetic ones, we need 
to be at a magnetic fixed point to start with. The only non-trivial such fixed 
point is at $T=T_c$, $H=0$; there we look for fixed points of the residual
coupling $J$. 

Before starting this search, 
let us recall that in two dimensions one can associate to a fixed point of the 
renormalization group a number $c$, called central charge, which grows with
the number of dynamical degrees of freedom \cite{BPZ,zamocth,cardycth}. A fixed
point (i.e. a conformal field theory) with central charge $c$ parameterized as 
\EQ
c=1-\frac{6}{m(m+1)}
\label{c}
\EN
contains scalar fields $\varphi_{r,s}$ (called primaries) with scaling 
dimension
\EQ
X_{r,s}=\frac{[(m+1)r-ms]^2-1}{2m(m+1)}\,,
\label{x}
\EN
together with infinitely many other less relevant fields (called descendants) 
whose dimensions exceed (\ref{x}) by integers. The field $\varphi_{1,1}$ with 
dimension $0$ is the identity $I$. For $m$ integer and larger than 1 the theory
admits a `minimal' realization in which the operator product expansion
closes on the primaries with $r$ and $s$ positive integers up to $m-1$ and
$m$, respectively, and their 
descendants \cite{BPZ}. Considering that within this set of primaries each 
dimension appears twice and excluding the identity, the number of independent 
and non-trivial primaries within a minimal model is $m(m-1)/2-1$.

Back to our problem, since no dynamical degrees of freedom are associated to 
the percolative properties, for all the fixed points with $T=T_c$, $H=0$ the 
central charge is that of the critical Ising model, i.e. $c=1/2$ ($m=3$).
Clearly, the decoupling point $J=0$ yields a trivial fixed point at which 
percolation
plays no role ($p_B=0$) and we deal with the pure magnetic fixed point of the 
Ising model. The latter is well known to be described by the minimal 
realization of the $m=3$ conformal theory, which indeed contains two relevant
scalar fields: the spin field $\sigma$ with dimension $X_\sigma=X_{1,2}=1/8$,
and the thermal field $\varepsilon$ with dimension $X_\varepsilon=
X_{1,3}=1$. It follows that at this fixed point the coupling $J$ is conjugated 
to an irrelevant field.

In order to progress, we need to recall that the dilute Potts model 
(\ref{potts}) (with $T$ which can be fixed to $T_c$ for the time being) admits 
two distinct lines of fixed points for $q<4$ \cite{NBRS}. The first one is the 
critical line of the undilute ($\Delta=+\infty$) Potts model \cite{Baxter}, 
while the second is a tricritical line arising at some $\Delta_c$ yielding a 
finite concentration of vacancies. Both lines are described by conformal 
theories with central charge (\ref{c}) and the following relations between $q$ 
and $m$ \cite{Nienhuis,DF}
\EQ
\sqrt{q}=2\sin\frac{\pi(t-1)}{2(t+1)}\,,\hspace{1.5cm}
m=\left\{
\begin{array}{l}
t\hspace{1.2cm}\mbox{for the critical line}\,,\\ 
\\
t+1\hspace{.6cm}\mbox{for the tricritical line}\,.
\end{array}
\right.  
\label{q}
\EN
The two fixed lines meet at $q=4$ ($m=\infty$, $c=1$) and the transition is
first order above this value. The Potts spin field has scaling dimension 
$X_s(q)$ which coincides with $X_{(m-1)/2,(m+1)/2}$ along the critical line, 
and with $X_{m/2,m/2}$ along the tricritical one. As for the fields invariant 
under the $S_q$ symmetry of color permutation, the leading scaling dimension 
$X_{t_1}(q)$ along the critical (tricritical) line is $X_{2,1}$ ($X_{1,2}$); 
since also the dilution is relevant at  tricriticality, a second relevant, 
$S_q$-invariant scaling field is present along the tricritical line and 
corresponds to $\varphi_{1,3}$, with dimension $X_{t_2}(q)=X_{1,3}$.

In the limit $q\to 1$ relevant for the percolative properties, the critical 
line gives $m=2$, $c=0$, in agreement with the fact that the undilute case
corresponds to random percolation, which carries no dynamical degrees of 
freedom. Hence, the critical exponents for random percolation are determined
by those of the $q\to 1$ pure Potts model. In particular, the cluster size 
exponent coincides with the Potts susceptibility exponent $\gamma$. This and 
the correlation length exponent $\nu$ are given by\footnote{We recall that the
values of the critical exponents depend on the direction in coupling space
along which the fixed point is approached. In (\ref{nu-gamma}) we take into
account that on the tricritical line there are two $S_q$-invariant relevant 
fields.}
\EQ
1/\nu=2-X_{t_i}\,,\hspace{1cm}\gamma/\nu=2-2X_s\,,
\label{nu-gamma}
\EN
and at the random percolation point are determined by the values $X_s(1)=5/48$ 
and $X_{t_1}(1)=5/4$ on the critical line. These values characterize the 
infinite temperature fixed point of Fig.~1. Notice that the minimal realization
of the $m=2$ fixed point contains only the identity: critical percolative 
properties are not described by minimal conformal field theories.

The limit $q\to 1$ along the tricritical line gives $m=3$, $c=1/2$. This is 
also expected because at $q=1$ there are no Potts degrees of freedom and the 
critical degrees of freedom are those of the lattice gas, which is an Ising 
model. As a consequence, within the model (\ref{potts}) this fixed point 
corresponds to $T=T_c$, $H=0$ and to some $J=J^*$. Contrary to the case $J=0$ 
discussed above, it is a non-trivial fixed point for correlated percolation.
The percolative exponents are determined here by the tricritical values 
$X_s(1)=5/96$, $X_{t_1}(1)=1/8$ and $X_{t_2}(1)=1$. Once again, the first of 
these dimensions does not belong to the minimal realization of the $m=3$ 
theory. 

Notice that the dimensions of the the two relevant $S_q$-invariant fields on 
the tricritical line coincide at $q=1$ with those of the Ising spin and thermal
fields, which are conjugated to $H$ and $T$, respectively. As a consequence, 
the remaining coupling $J$ in (\ref{potts}) is irrelevant at $J^*$. Since it
was irrelevant also at $J=0$, consistence of the renormalization group flows
requires a third, intermediate fixed point where $J$ is relevant (Fig.~3).
This was located in \cite{CK} at $J=2/T_c$ as a consequence of the identity
\EQ
\left.-{\cal H}_q\right|_{J=2/T}=\frac2T\sum_{\langle ij\rangle}
(\delta_{\nu_i,\nu_j}-1)+(\ln q-2H)
\sum_i\delta_{\nu_i,0}\,,\hspace{1cm}\nu_i=0,1,\ldots,q\,.
\label{q+1}
\EN
This equation is easily checked using $t_i=\delta_{\sigma_i,+}$, 
$\delta_{\sigma_i,\sigma_j}=
\delta_{\sigma_i,+}\delta_{\sigma_j,+}+\delta_{\sigma_i,-}\delta_{\sigma_j,-}$ 
and $\delta_{\sigma_i,-}\delta_{\sigma_j,-}+\delta_{s_i,s_j}\delta_{\sigma_i,+}
\delta_{\sigma_j,+}=\delta_{\mu_i,\mu_j}$, where $\mu_i$ is a site variable 
which takes the value $s_i$ if $\sigma_i=1$, and $0$ if $\sigma_i=-1$. 
Substitution into the original form of the Hamiltonian gives (\ref{q+1})
with $\mu_i$ instead of $\nu_i$ and without the $\ln q$ term. The latter arises
because the $q$ ways of coloring a site with spin down produce a factor 
$q^{N_v}$ when switching from the partition sum over $\{\mu_i\}$ to that over
$\{\nu_i\}$. The Hamiltonian (\ref{q+1}) is that of a $(q+1)$-state Potts 
model and is critical for $2H=\ln q$ and some $q$-dependent value of $T$. As
$q\to 1$ we obtain an Ising model with critical point at $H=0$ and $T=T_c$.

\begin{figure}
\centerline{
\includegraphics[width=7cm]{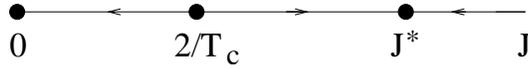}}
\caption{Renormalization group flows in the coupling $J$ for the Hamiltonian
(\ref{potts}) in two dimensions with $q\to 1$, $T=T_c$, $H=0$. While $J=0$ 
is just the Ising magnetic fixed point, $2/T_c$ and $J^*$ are also percolation 
points for Ising droplets and Ising clusters, respectively.
}
\end{figure}

It is clear from (\ref{q+1}) that $J=2/T$ is a special case of (\ref{potts}) in
which the lattice gas variable and the Potts spin are treated symmetrically,
so that they have the same scaling dimension at the fixed point $J=2/T_c$:
$X_s=X_\sigma=1/8$; similarly, $X_{t_1}=X_{\varepsilon}=1$. This means that the
KF clusters with $p_B=1-e^{-2/T}$ percolate at $T_c$ with exponents $\gamma$
and $\nu$ which coincide with those of the magnetic susceptibility and 
correlation length in the Ising model. Since these are requirements of the 
droplet model \cite{Fisherdroplet} meant to describe magnetic transitions in a 
cluster language, the KF clusters with $p_B=1-e^{-2/T}$ are called
Coniglio-Klein Ising droplets \cite{CK}.

Concerning Ising clusters, the renormalization group pattern of Fig.~3 finally
leads to the conclusion that their critical exponents at $T=T_c$, $H=0$ are 
those of the fixed point $J^*$, onto which $J=\infty$ ($p_B=1$) renormalizes
at large distances \cite{CK}. The mean cluster size exponent in the limit
$T\to T_c^-$, $H=0$ was first evaluated by series expansions in \cite{SG}, 
with the result $\gamma\approx 1.91$, quite close to the exact one
$\gamma=91/48=1.895..$ coming from $X_s=5/96$ and $\nu=1$. This exact value
was identified in \cite{SV}.

We finish this section recalling that at the percolation point clusters behave 
as fractals, i.e. their size grows with the linear extension $L$ as $L^D$, with
a fractal dimension $D$ smaller than the space dimensionality $d$ \cite{SA}. 
The fractal dimension is easily determined considering the number $S_\infty(L)$
of sites belonging to the incipient infinite cluster inside a box of side $L$. 
This goes like $PL^d$, where $P$ is the density of the infinite cluster, namely
the percolative order parameter, which goes like the Potts magnetization per 
site, i.e. like $L^{-X_s}$.
Hence $D=d-X_s$. In $d=2$ the values of $X_s$ given above determine the 
fractal dimensions $91/48=1.89..$ for random percolation, $187/96=1.94..$ for 
Ising clusters, and $15/8=1.87..$ for Ising droplets, which are less dense than
Ising clusters, as expected.

\section{Field theory of the scaling limit}
The scaling region of lattice models around second order phase transition 
points can be described in a continuous, field theoretical framework. With 
the notation for the fields introduced in the previous section, the scaling 
limit of the Hamiltonian (\ref{ising}) in $d=2$ is described by the Ising 
field theory (\cite{report} for a review) with action
\EQ
{\cal A}_{Ising}={\cal A}_{CFT}^{Ising}-\tau\int d^2x\,\varepsilon(x)-h\int 
d^2x\,\sigma(x)\,,
\label{isingft}
\EN
where ${\cal A}_{CFT}^{Ising}$ is the action of the conformal field theory with
central charge $c=1/2$, and $\tau\sim T-T_c$, $h\sim H$ as the critical point
is approached. The action (\ref{isingft}) encodes all the universal features
of the ferromagnetic phase transition. The universal percolative properties of
the Ising model are instead contained in the scaling limits of the dilute Potts
Hamiltonian (\ref{potts}) with $q\to 1$. It follows from the discussion of the
previous section that, as far as Ising clusters are concerned, the 
renormalization group trajectories which matter are those originating from 
the line of tricritical fixed points. These are described by the field theory
\EQ
{\cal A}_{q}={\cal A}_{CFT}^{tricr}-g\int d^2x\,\varphi_{1,3}(x)-\lambda
\int d^2x\,\varphi_{1,2}(x)\,,
\label{pottsft}
\EN
where ${\cal A}_{CFT}^{tricr}$ is the action of the conformal theory with 
central charge (\ref{c}), related to $q$ by (\ref{q}) with $t=m-1$, and, as we
saw, $\varphi_{1,2}$ and $\varphi_{1,3}$ are the two relevant $S_q$-invariant 
fields at tricriticality. 

\begin{figure}
\centerline{
\includegraphics[width=6cm]{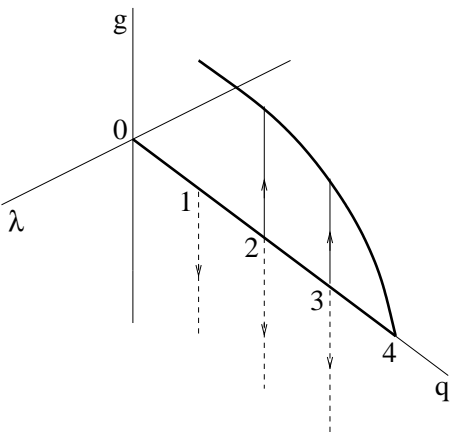}}
\caption{Phase diagram of the field theory (\ref{pottsft}) describing the 
scaling dilute Potts model. The ferromagnetic phase transition surface 
corresponds to $\lambda=0$. On this surface, the trajectories originating from 
the tricritical line are massive for $g<0$ (first order transition); they are 
massles and flow into the critical line for $g>0$ (second order transition).
}
\end{figure}

Let us first discuss some general features of the field theory (\ref{pottsft})
for $1<q\leq 4$. It follows from general results \cite{Taniguchi} that 
the theory is integrable when at least one of the two 
couplings $g$ and $\lambda$ vanishes. Integrability allows to establish that
the critical surface separating magnetically ordered
and disordered regions of the scaling dilute Potts model corresponds to 
$\lambda=0$ (Fig.~4): for $\lambda=0$, $g<0$, (\ref{pottsft}) describes the 
first order part of the transition surface on which the $q$ ordered ground 
states are degenerate with the disordered one \cite{dilute}; for $\lambda=0$, 
$g>0$, instead, the theory is massless \cite{zamocth,FSZ} and describes the 
second order part of the transition surface, spanned by the trajectories 
flowing from the tricritical to the critical line of fixed points.

For $\lambda<0$, the theory (\ref{pottsft}) possesses a unique vacuum $|\Omega
\rangle$ corresponding to the disordered ground state; for $\lambda>0$ the
degenerate vacua $|\Omega_i\rangle$, $i=1,\ldots,q$, correspond to the $q$
ordered ground states. 
The order parameter of the ferromagnetic transition is the Potts magnetization
per site $M_q$, i.e. the expectation value of the field $\delta_{s(x),j}-1/q$
taken over $|\Omega\rangle$ for $\lambda<0$, and over $|\Omega_j\rangle$ for 
$\lambda>0$. At the first order transition the $q+1$ vacua are degenerate,
i.e. $\langle\varphi_{1,3}\rangle$ is the same on all vacua. The expectation
value $U_q\equiv\langle\varphi_{1,2}\rangle$, like $M_q$, is discontinuous 
across the first order transition surface, its discontinuity corresponding to 
the latent heat \cite{dilute}. Both
$M_q$ and $U_q$ vanish on the second order transition surface (Fig.~5).

We are now ready to discuss the limit we are actually interested in, 
$q\to 1^+$, which we know a priori is peculiar. Indeed, it follows
from the scaling dimensions given in the previous section that, as expected,
${\cal A}_1={\cal A}_{Ising}$, with $g=\tau$ and $\lambda=h$. Since 
${\cal A}_{Ising}$ is massive\footnote{Field theoretically, due to the 
reflection positivity of (\ref{isingft}), this conclusion follows from 
Zamolodchikov's $c$-theorem \cite{zamocth}.} for any value of the couplings 
away from $\tau=h=0$, the massless trajectories of (\ref{pottsft}) for 
$\lambda=0$, $g>0$ have to become massive as $q\to 1$. Remarkably, this
phenomenon can be described analytically exploiting the integrability of 
(\ref{pottsft}) with $\lambda=0$.

\begin{figure}
\centerline{
\includegraphics[width=12cm]{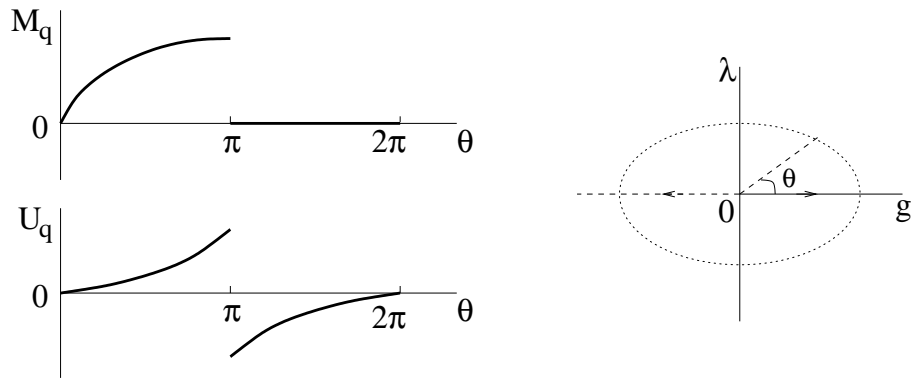}}
\caption{Qualitative behavior (along the dotted path on the right) of the 
magnetic order parameter $M_q$ and of $U_q=\langle\varphi_{1,2}\rangle$ in the 
scaling dilute Potts model (\ref{pottsft}). For $q=1$ this model becomes the
Ising field theory (\ref{isingft}) with $g=\tau$ and $\lambda=h$; $M_1$ and 
$U_1$ become the Ising percolative and magnetic order parameter, respectively.
}
\end{figure}

Let us recall that use of (non-conformal) integrability in two-dimensional 
field theory 
requires switching to a particle language. Indeed, integrable field theories
are solved by exact determination of the $S$-matrix of the associated
scattering theory in ($1+1$)-dimensional space-time \cite{ZZ}. In the ordered 
phase $\lambda>0$, the elementary excitations of (\ref{pottsft}) on which the 
scattering theory is built are the $q(q-1)/2$ kinks which interpolate between 
the $q$ degenerate vacua, together with their antikinks; due to $S_q$ symmetry 
all these kinks have the same mass \cite{CZ}. In the disordered phase 
$\lambda<0$, instead,
the elementary excitations are ordinary particles $A_k$, $k=1,\ldots,q-1$, 
forming a degenerate multiplet in which the antiparticle $\bar{A}_k$ coincides
with $A_{q-k}$ \cite{qconf}. The mass of all these excitations goes to zero
when the second order transition surface is approached ($\lambda\to 0$, $g>0$).
Since the $q$ vacua of the spontaneously broken phase coalesce in this limit,
the elementary excitations on the transition surface are to be identified with
the massless limit of the particles $A_k$, which, in ($1+1$) dimensions, are
right/left movers with energy $p^0=(\mu/2)e^{\pm\theta}$ and momentum $p^1=\pm 
p^0$, $\mu$ being a mass scale and $\theta$ a rapidity parameter.

While the number of these massless degrees of freedom clearly vanishes as
$q\to 1$, the emergence of a massive particle in the same limit follows from 
the study of the right-left scattering amplitudes\footnote{See 
\cite{Alioshaflow,ZZmassless} for the theory of integrable massless 
scattering.}. The latter are known exactly \cite{FSZ} within a particle basis 
which is not the Potts basis we are discussing\footnote{See \cite{FR} for a 
discussion on $S_q$-invariant scattering theories and change of particle 
basis.}. This is, however, immaterial as far as the analytic properties we are 
interested in are concerned (see e.g. \cite{RSOS}), and the only think we need 
to know is that the poles of the amplitudes are determined by the factor
\cite{FSZ}
\EQ
\frac{1}{\cosh\rho(i\pi-\theta)}\exp\left[-i\int_0^\infty\frac{dx}{x}
\frac{\sinh\frac{x}{2}}{\sinh\frac{x}{2\rho}\cosh\frac{x}{2}}\sin
\frac{\theta x}{\pi}\right]\,, 
\label{poles}
\EN
where $\rho=1/(m-1)$  and $\theta$ determines the square of the center of mass 
energy as $s=\mu^2e^\theta$. As usual, the scattering amplitudes have a 
multi-sheet structure in the complex $s$-plane. The `physical sheet' on this 
plane corresponds to $\mbox{Im}\,\theta\in(0,\pi)$, and the `second sheet' to 
$\mbox{Im}\,\theta\in(0,-\pi)$. Writing the exponential part of (\ref{poles})
as $S_{-1/2}(\theta)/S_{1/2}(\theta)$, with
\EQ
S_{\gamma}(\theta)=\prod_{n=0}^\infty\frac{
\Gamma\left(\frac12+\left(2n+\frac32-\gamma\right)\rho-
\frac{\rho\theta}{i\pi}\right)
\Gamma\left(\frac12+\left(2n+\frac12-\gamma\right)\rho+
\frac{\rho\theta}{i\pi}\right)}{
\Gamma\left(\frac12+\left(2n+\frac32-\gamma\right)\rho+
\frac{\rho\theta}{i\pi}\right)
\Gamma\left(\frac12+\left(2n+\frac12-\gamma\right)\rho-
\frac{\rho\theta}{i\pi}\right)
}\,,
\EN
one can see that for $q>1$ (i.e. $m>3$) (\ref{poles}) has no poles 
on the physical sheet, and possesses a single pole, coming from the $1/\cosh$ 
prefactor, at $\theta_0=-i\pi(m-3)/2$, which is on the second sheet for 
$m\in(3,5)$. This pole at $s_0=\mu^2e^{\theta_0}$ on the second sheet (Fig.~6)
corresponds to a resonant particle $A$ in the right-left
scattering channel. When $q\to 1^+$, the pole approaches the 
physical sheet and gives a narrow resonance with square mass $\mbox{Re}\,s_0
\simeq\mu^2$ and inverse lifetime $\mbox{Im}\,s_0/\mu\propto(q-1)\mu$, which 
coexists with the $q-1$ massless particles. At $q=1$, no massless particles are
left and $A$ provides the {\it stable} massive particle of the Ising field 
theory (\ref{isingft}) with $h=0$, $\tau>0$. 

\begin{figure}
\centerline{
\includegraphics[width=5.5cm]{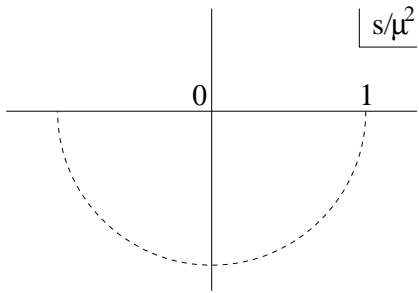}}
\caption{Location $s_0$ of the resonant pole on the second sheet of the 
$s$-plane. The pole describes an anticlockwise motion along the dashed path as 
$m$ decreases from $5$ to $3$: $s_0$ is purely imaginary at $m=4$ ($q=2$) and 
real and positive at $m=3$ ($q=1$), where the associated particle becomes 
stable.
}
\end{figure}

Since the number of components of $S_q$-multiplets vanishes as $q\to 1$, the
particle $A$ we are talking about must be an $S_q$-singlet, i.e. arises in 
the right-left scattering channels $A_kA_{q-k}$. This prevents the correlators
of $S_q$-invariant fields from vanishing as $q\to 1$. Indeed, let us denote by 
$\phi$ such a field (scalar, for simplicity) and consider the spectral 
decomposition of its two-point function $G(x)=\langle\phi(x)\phi(0)\rangle$ 
over asymptotic particle states, in the limit $q\to 1$. The contribution to 
$G(x)$ with the lowest number of intermediate particles is
\EQ
\lim_{q\to 1}\,
(q-1)\int d\theta_1d\theta_2\left|\langle\Omega|\phi(0)|A_k(\theta_1)A_{q-k}
(\theta_2)\rangle\right|^2e^{-|x|E_{2,0}(\theta_1,\theta_2)}\,,
\label{2part}
\EN
where the prefactor $q-1$ comes from the summation over $k$, and $E_{2,0}
(\theta_1,\theta_2)=\mu(e^{\theta_1}+e^{-\theta_2})/2$ is the energy of a 
right-left pair. Right-right and left-left pairs do not contribute to 
the limit because only the
right-left matrix element inherits from the right-left scattering amplitude
the resonance pole at $\theta_1-\theta_2=\theta_0$, with $\theta_0\propto-i(q-
1)$ for $q\to 1$. Hence, calling $R_\phi$ the residue of the matrix element on 
this pole, (\ref{2part}) can be written as\footnote{By relativistic invariance,
the two-particle matrix element of a scalar field depends on the rapidity
difference, and the one-particle matrix element is rapidity-independent.}
\EQ
\lim_{q\to 1}\,(q-1)\int d\beta d\theta\,\frac{|R_\phi|^2}{(\theta-\theta_0)
(\theta+\theta_0)}\,e^{-|x|E_{2,0}((\beta+\theta)/2,(\beta-\theta)/2)}
\propto\int d\beta\,e^{-|x|E_{1,\mu}(\beta)}\,,
\EN
where $E_{1,\mu}(\beta)=\mu\cosh(\beta/2)$ is the energy of a particle with
mass $\mu$. We see in this way how correlations mediated by massless particles 
on the second order surface are replaced by correlations mediated by the 
particle $A$ along the massive trajectory at $q=1$ (Fig.~7). Fields which are 
not $S_q$-invariant do not couple to $A$ and have zero correlations (i.e. are 
absent) at $q=1$ (as required by $S_1=I$). 
Recalling that the $S_q$-invariant fields along the tricritical Potts line are
$I$, $\varphi_{1,2}$, $\varphi_{1,3}$ and their descendants, which become $I$, 
$\sigma$, $\varepsilon$ and their descendants at $q=1$, the mechanism we 
illustrated explains in particular how the theory becomes minimal for this 
value of $q$.

Summarizing, the percolation line $H_0(T)$ of Fig.~1 is mapped in the scaling
limit onto the case $h=0$ of the field theory (\ref{isingft}). More precisely,
as expected from (\ref{free-energy}), this field theory describes only the 
magnetic properties of the Ising model; the universal percolative properties of
Ising clusters are described by the `embedding' of (\ref{isingft}) into the
dilute Potts field theory (\ref{pottsft}) with $q\to 1$. In this limit, the 
expectation values $M_q$ and $U_q$ shown in Fig.~5 become the Ising percolative
and magnetic order parameter, respectively. While for $\tau<0$ both transition
are first order, for $\tau>0$ there is only a continuous percolative 
transition, a circumstance which is explained analytically by the evolution of
the resonance pole discussed above. 

Notice that for the transition at $\tau<0$ the situation is 
simpler. On the first order part of the transition surface of (\ref{pottsft})
the $q$ ordered vacua $|\Omega_i\rangle$ are degenerate with the disordered 
one $|\Omega\rangle$. The elementary excitations are $q$ kinks interpolating 
from $|\Omega\rangle$ to $|\Omega_i\rangle$, together with their antikinks 
\cite{dilute}. When $q\to 1$ one recovers straightforwardly the single kink
of the Ising model.

\begin{figure}
\centerline{
\includegraphics[width=8.5cm]{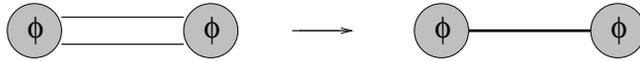}}
\caption{Correlations mediated by two massless particles at $q>1$ are 
mediated by one massive particle at $q=1$. Fields $\phi$ which are not 
$S_q$-invariant do not couple to the massive particle and are absent at $q=1$.
}
\end{figure}

\section{Universal scaling limit of the Kert\'esz line}
We saw in section~3 that for $H=0$ the Coniglio-Klein droplets, 
i.e. the KF clusters 
with $p_B=1-e^{-2/T}$, percolate at $T_c$ (also in $d=3$) with critical 
exponents coinciding with the magnetic exponents. Also, contrary to the case of
Ising clusters, an infinite droplet above $T_c$ in zero field is excluded 
by the vanishing of $p_B$ at infinite termperature. It is easy to see, however,
that when $H\neq 0$ the critical properties of the magnetic and percolative
degrees of freedom no longer coincide. Indeed, for $H=+\infty$ all the spins
are up and we are left with a random bond percolation problem with occupation
probability $p_B$, which is critical for some value $p_B^c$. This means that 
there is a line $T_K(H)$, called the Kert\'esz line \cite{Kertesz}, going from 
$T_K(0)=T_c$ to $T_K(+\infty)=-2/\ln(1-p_B^c)$, which is a percolation line for
the Coniglio-Klein droplets (Fig.~1). 

The universal features of these droplets are described by the scaling limit of 
the Hamiltonian (\ref{q+1}), namely, in $d=2$, by the field theory
\EQ
{\cal A}_{droplets}={\cal A}_{CFT}^{(q+1)}-\tau_q\int d^2x\,\varphi_{2,1}(x)+
2h_q\int d^2x\,\delta_{\nu(x),0}\,,\hspace{1cm}\nu(x)=0,1,\ldots,q\,,
\label{droplet}
\EN
where ${\cal A}_{CFT}^{(q+1)}$ accounts for the critical line of a 
$(q+1)$-state Potts model, $\tau_q$ measures the deviation from the
critical Potts temperature, and $h_q$ is a magnetic field pointing in the 
$\nu=0$ direction. Since in two dimensions the Potts transition is second order
as long as the number of states does not exceed $4$, the above action is 
intended for $q\leq 3$. For $q=1$, (\ref{droplet}) with $\tau_1=\tau$ and 
$h_1=h$ gives back the Ising field theory (\ref{isingft}).

The renormalization group trajectories flowing out of the fixed point at 
$\tau_q=h_q=0$ for $h_q\geq 0$ are labelled by the dimensionless parameter
\EQ
\eta_q=\tau_q/h_q^{(2-X_{2,1})/(2-X_s)}\,,
\label{etaq}
\EN
where the thermal and magnetic scaling dimensions $X_{2,1}$ and $X_s$ are
those given in section~3 for the Potts critical line, up to the 
substitution $q\to q+1$; $\eta_1\equiv\eta=\tau/h^{8/15}$ labels the Ising
trajectories.
For $h_q=+\infty$ the state $\nu=0$ is forbidden and one is left with a 
$q$-state Potts model. This means that for $1<q\leq 3$ there is a critical 
trajectory $\eta_q^c$ which flows from the $(q+1)$- to the $q$-state Potts 
fixed point. The $S_q$-invariance of (\ref{droplet}) is spontaneously broken 
for $\eta_q<\eta_q^c$. Since the Potts Curie temperature increases as $q$ 
decreases,
$\eta_q^c$ is expected to be positive\footnote{The field theory (\ref{droplet})
has been considered in \cite{qconf} with the purpose of describing the 
qualitative evolution of the particle spectrum in parameter space. The actual 
value of $\eta_q^c$ was not essential there and was naively identified with 
zero.}.  

The massless trajectories $\eta_q^c$ span, as a function of $q$, a critical 
surface which plays for the droplets exactly the same role the second order
surface considered in the previous section played for the Ising clusters. In 
the present case, however, while it can be seen that the number of massless 
particles indeed vanishes as $q\to 1$ \cite{qconf}, lack of integrability does
not allow to follow analytically the origin of the mass gap at $q=1$. The 
most likely mechanism remains the one seen in the previous section, namely at 
least one  
neutral massive resonance which becomes stable at $q=1$. The possibility that
this neutral particle is stable already for $q>1$ cannot be excluded, but 
requires a mechanism which prevents the decay into the massless excitations.

\begin{figure}
\centerline{
\includegraphics[width=6cm]{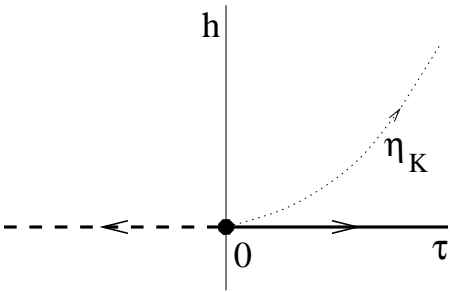}}
\caption{Scaling limit of Fig.~1 within the parameter space of the Ising field 
theory (\ref{isingft}). The percolative order parameter is non-zero in the 
whole upper half-plane for Ising clusters, and to the left of the Kert\'esz 
trajectory $\eta_K$ for Coniglio-Klein Ising droplets.
}
\end{figure}

Either way, it follows from the same line of arguments developed for the 
clusters that the value $\eta_1^c\equiv\eta_K$ of the Ising parameter $\eta$ 
determines the universal scaling limit of the Kert\'esz line (Fig.~8). An 
infinite Coniglio-Klein droplet exists in the sector of the $\tau$-$h$ plane 
with $h>0$, 
$\eta<\eta_K$. Percolative critical exponents measured along the trajectory 
$\eta_K$ (as along the trajectory $\eta=+\infty$ for clusters) are those of the
percolative infrared fixed point, i.e. the random percolation fixed point
with central charge $c=0$.

Numerical data for the Kert\'esz line of the square lattice Ising model in the
vicinity of the magnetic critical point are given in \cite{FS}. Using them,
together with the known relations between lattice and continuum parameters 
(see e.g. \cite{CGR}), we find\footnote{This and the subsequent values of 
$\eta$ refer to the normalization 
\EQ
\lim_{|x|\to 0}|x|^{1/4}\langle\sigma(x)\sigma(0)\rangle=
\lim_{|x|\to 0}|x|^{2}\langle\varepsilon(x)\varepsilon(0)\rangle=1
\label{cftnorm}
\EN
of the fields in (\ref{isingft}).} $\eta_K\simeq 0.12$. We 
recall that, as a result of a series of theoretical and numerical studies (see
\cite{report} for references), very much is known about the $\eta$-dependence 
of the particle spectrum of the field theory (\ref{isingft}). In particular,
the theory possesses a single stable particle for $\eta>\eta_{(2)}$, and 
two stable particles for $\eta_{(2)}>\eta>\eta_{(3)}$, with $\eta_{(2)}
\simeq 0.33$ and $\eta_{(3)}\simeq 0.022$ \cite{FZ}. It follows that $\eta_K$
falls inside the second region, so that, if the resonance scenario for
the production of the mass gap as $q\to 1$ along the critical surface applies 
also to this case, two particles need to become simultaneously stable at $q=1$.

\section{Conclusion}
In this paper we showed explicitly the non-perturbative analytic mechanism 
through
which clusters of up spins in the two-dimensional Ising model undergo a 
second order percolation transition at values of temperature $T$ and magnetic 
field $H$ for which there are no 
thermodynamic singularities. The Kasteleyn-Fortuin formulation of percolation 
leads to consider a dilute $q$-state Potts model
in which the Ising degrees of freedom are associated to the dilution and the 
Potts spins are auxiliary site variables. In the three-dimensional parameter
space of $T$, $H$ and $q$ there is for $q>1$ a surface of second order phase
transition which, in the scaling limit, corresponds to an integrable field theory.
This allows to show that the $q-1$ massless particles on the surface produce 
in their scattering a massive resonance whose lifetime is proportional to 
$1/(q-1)$. When the limit $q\to 1$, needed to get rid of the auxiliary Potts 
variables, is taken, the percolative correlations, which are determined at first 
order in $q-1$, are mediated by the massless particles, giving rise to the 
second order transition line. The thermodynamic observables, instead, are 
determined at $q$ strictly equal to 1, where no massless particle is left and
the correlations are mediated by the massive particle which becomes 
stable. The fact that the resonant particle is a singlet under the Potts 
permutational symmetry $S_q$ leads to a smooth transition from the massless to 
the massive regime for the thermodynamic observables as $q\to 1$. 

We have analyzed in the same framework also the Coniglio-Klein droplets,
which are obtained from the Ising clusters by a partial, temperature-dependent
depletion and produce at the Curie point percolative
exponents which coincide with the magnetic ones. The corresponding field
theory in the Kasteleyn-Fortuin representation describes the scaling limit of 
a $(q+1)$-state Potts model with a magnetic field which reduces the symmetry 
to $S_q$. Again, for $q>1$ there is a massless surface bounded at 
$q=1$ by a massive trajectory which corresponds
to the universal scaling limit of the second order percolation line for the 
droplets (the Kert\'esz line). In this case the massless surface is not 
integrable, but a mechanism analogous to that observed for the clusters is 
likely to account for mass generation at $q=1$. 

On the lattice, the second order percolative line for the clusters (droplets)
goes from the Curie point to a finite value of the magnetic field (temperature)
at infinite temperature (magnetic field). The asymptotic value is determined 
by the critical probability of random site (bond) percolation on the given
lattice, and is non-universal. Our field theoretical description of the scaling
limit retains only the universal features. The second order critical lines
correspond to renormalization group trajectories flowing from the correlated
percolation fixed point, with central charge $c=1/2$ and different dimensions
of the percolative order parameter for clusters and droplets, to the random 
percolation fixed point with $c=0$. These universal trajectories are identified
by two values of the Ising field theory parameter 
$\eta=\tau/h^{8/15}$: $\eta=+\infty$ for the clusters and 
$\eta\simeq 0.12$ for the droplets. The latter value is obtained from existing 
numerical data and implies that the Kert\'esz trajectory falls in the 
sector of the Ising field theory with two stable massive particles.

\vspace{1cm}\textbf{Acknowledgments.} I am grateful to M. Caselle and P. Grinza
for conversations on the Kert\'esz line. 
Work supported in part by the ESF grant INSTANS and by the MIUR grant ``Fisica
statistica dei sistemi fortemente correlati all'equilibrio e fuori 
equilibrio: risultati esatti e metodi di teoria dei campi''.

\end{document}